\documentstyle{amsppt}

\TagsOnRight
\newdimen\theight
\define\lab#1{\vadjust{\setbox0=\hbox{\tenpoint%
            \quad${}^{\dsize\checkmark}{}^{\text{#1}}$}%
            \theight=\ht0
            \advance\theight by\dp0\advance\theight by \lineskip
            \kern-\theight\vbox to\theight{\rightline{\rlap{\box0}}%
            \vss}%
            }}%
\mathsurround=1.2pt
\define\<{\leavevmode\kern-\mathsurround}

\define\>{\hskip1pt}
\redefine\le{\leqslant}
\redefine\ge{\geqslant}
\define\){)\hskip1pt}
\redefine\({\hskip1pt(}
\define\op{\operatorname}
\define\wt{\widetilde}
\define\wh{\widehat}
\define\e{\varepsilon}
\define\ov{\overline}
\redefine\pa{\partial}

\topmatter
\title Exactly Solvable Two-Dimensional Schr\"odinger Operators \\
and Laplace Transformations
\endtitle
\rightheadtext{Exactly solvable two-dimensional Schr\"odinger operators}
\author S.~P.~Novikov and A.~P.~Veselov
\endauthor
\affil S.P.Novikov:
University of Maryland at College Park,  Department of Mathematics
and IPST,
College Park MD 20742-2431 USA 
and Landau Institute for Theoretical Physics,
Moscow 117940, Kosygin street 2, Russia. E-mail adress novikov\@ipst.umd.edu
A.P.Veselov:
Department of Mathematical Sciences, Loughborough University,
Loughborough, Leicestershire, LE11 3TU, UK and
Department of Mathematics and Mechanics, Moscow State University,
119899 Moscow Russia. E-mail adress a.p.veselov\@lboro.ac.uk
\endaffil
\translator A. SOSSINSKY
\endtranslator
\endtopmatter

\document

\head\S1.\enspace Solitons and the Schr\"odinger operator \endhead

The modern theory of exactly solvable one-dimensional and
two-dimensional Schr\"odinger operators
$$
\gather
L=-\pa_x^2+u(x)\quad(n=1),\\
L=(\ov{\pa}+B(x,y))(\pa+A(x,y))+2V(x,y),\quad\pa=\pa_x-i\pa_y,\;z=x+iy
\quad(n=2)
\endgather
$$
cannot be separated from the theory of spatial one- and
two-dimensional integrable nonlinear systems of soliton theory.
This is especially so in the rapidly decaying and periodic case (when
the coefficients of the operators are rapidly decaying and periodic
functions of the space variables).

\subhead I. One-dimensional case \rm(\<$n=1$)\endsubhead
In the one-dimensional case, it follows from the papers
\cite{GGKM, L}, where the famous KdV equation
$$
u_t=6uu_x-u_{xxx}
$$
was solved in the rapidly decaying case that it can be viewed as a
kind of ``symmetry'' of the spectral theory of the class of operators
$L=-\pa_x^2+u(x)$ acting in the Hilbert space $\Cal L_2(\Bbb R)$.
This means that the KdV equation can be written in the form
$$
L_t=AL-LA=[A,L],
$$
where $A$ is the linear differential operator
$A=-4\pa_x^3+3\(u\pa_x+\pa_xu)$. Therefore, for
example, the eigenvalues $L\psi=\lambda_i\psi$ of the operator $L$ on
any translation-invariant class of functions of $x$ on the line $\Bbb R$
turn out to be integrals of the KdV equation: $d\lambda_i/dt=0$.
Using the previously solved inverse scattering problem, KdV
was solved in \cite{GGKM} for rapidly decaying functions. Beginning
with S.~P.~Novikov's paper \cite{N74}, in the articles \cite{DN74, D75,
IM, L75, MV} the solution of the periodic problems for nonlinear KdV
from soliton theory was obtained simultaneously with that of the
inverse problems for the spectral theory of the linear operator $L$,
which had not been previously solved in the periodic case (see the
surveys \cite{DMN, Kr77, DKN2} and the book \cite{Sol}). The higher
analogs of KdV yield an infinite-dimensional commutative group of
such ``spectral symmetries'' for the operator $L$. The
finite-dimensional orbits of this group generate the famous
rapidly decaying and quasiperiodic (``multi-soliton and
finite-zone'') solutions of KdV as well as the remarkable exactly
solvable Schr\"odinger operators on the line in $\Cal L_2(\Bbb R)$.
Here the hyperelliptic $\Theta$-functions of Riemann surfaces make
their appearance, together with a series of other now widely known
formulas and results. Such are the consequences of continuous
spectral symmetry.

A few years ago some interesting advances were made \cite{W, SY,
S1} in the application of discrete symmetries, such as sequences of
the well-known B\"acklund--Darboux transformations, to the spectral
theory of the operator $L=-\partial_x^2+u(x)$. 
Probably the most significant were
obtained by A.~B.~Shabat and A.~P.~Veselov \cite{VS1}.

\subhead II. Two-dimensional topologically trivial case \rm(\<$n=2$)
\endsubhead
Beginning with the papers \cite{M, DKN1} in 1976, as the basis of the
two-dimensional analog of the theory of integrable systems associated
with the two-dimensional Schr\"odinger operator, it is customary to
take a representation of the form
$$
L_t=L\wh A+\wh B L,
$$
where $L$ is the two-dimensional Schr\"odinger operator (see
above), $\wh A$ and $\wh B$ are differential operators on
the plane $(x,y)$. Here the data of the inverse problem for the
spectral theory of the linear operator $L$ are taken from one level
only, namely from the set of solutions of $L\psi=0$ under some
``boundary'' condition or other on~$\psi$. For the operator $L$ with
doubly-periodic coefficients one takes the Bloch waves
$$
\wh T_\alpha\psi=\exp\>\{ip_\alpha T_\alpha\}\psi,\quad
L\psi=0,\qquad\alpha=1,2.
$$
Here $T_\alpha$ are the periods, $\alpha=1,2$, while $\wh
T_\alpha$ are the translation operators by the period $T_\alpha$, and
$p_\alpha$ is the quasi-momentum.
The set of all complex Bloch waves
of zero level  $L\psi=0$ constitute a one-dimensional complex
manifold $\Gamma$: $\psi(P,z,\ov z)$, $P\in\Gamma$ (the Fermi
curve). Without going into the details of the analytic properties of
the function $\psi$, first clarified in \cite{DKN1}, let us note the
following: exactly solvable topologically trivial two-dimensional
Schr\"odinger operators correspond to the case in which one Fermi
curve $\Gamma$ is of finite genus. The selection of the inverse
problem data corresponding to the purely potential operators
$$
L=\pa\ov{\pa}+V(x,y)
$$
was carried out in \cite{NV1} (see the survey \cite{NV2}). The
simplest two-dimensional system of soliton theory, generalizing
KdV, is of the form
$$
\frac{\pa V}{\pa t}=(\pa^3+\ov{\pa}{}^3\)V+
\pa{(uV)}+\ov{\pa}(\ov{u}V),\qquad \ov{\pa}u=3\pa V
$$
and can be represented as
$$
\gather
\frac{\pa L}{\pa t}=[L,\wh B]+fL,\\
\wh B=({\pa}^3+\ov{\pa}{}^3)
+(u{\pa}+\ov u\ov{\pa}),\qquad f={\pa}u+\ov{\pa}\>\ov u.
\endgather
$$

There is an entire infinite-dimensional commutative group of systems
of this form (the {\it higher analogs\/} of two-dimensional KdV or {\it
Novikov--Veselov hierarchy\/}).

Such is the two-dimensional analog of the continuous spectral
symmetry of the class of Schr\"odinger operators. In contrast to
the one-dimensional case, it acts only on the eigenfunctions of one
level  $L\psi=0$.

Already at the end of 18th century, Laplace (see \cite{L}) proposed
substitutions acting on the solutions of the equation $L\psi=0$ and
transforming them into solutions of a new Schr\"odinger equation
$\wt L\wt\psi=0$. We call them ``Laplace
transformations''. In the present paper we use the Laplace
transformations as a kind of ``discrete symmetry'' for the spectral
theory of the Schr\"odinger operator $L$ corresponding to one level
 $L\psi=0$, i.e., being in good agreement with the ideology of
two-dimensional soliton theory described above.

More precisely, we shall construct new classes of two-dimensional
Schr\"odinger operators that are exactly solvable at one or several
energy levels also in the {\it topologically nontrivial case}, a case
about which little is known (practically nothing, except the papers
\cite{AC, DN80} and the survey of one of the authors \cite{N83}). It
is precisely this last part of the present paper that we regard as
its main result. Part of the results of the present paper was
already announced in \cite{NV3}.

An alternative approach, based not only on Laplace trancformations,
but also on additional symmetries, is being developed by A.B.Shabat
and A.P.Veselov in \cite{S2} , \cite{SV2} (in preparation). 
Its relation with 
spectral theory of topologically
nontrivial operators (see below) is totally unclear.

\head\S2.\enspace Two-dimensional Schr\"odinger operator in a magnetic
field. Topologically nontrivial operators. The Pauli operator
\endhead

The general two-dimensional Schr\"odinger operator on the Euclidean
plane $\Bbb R^2$ has the form
$$
L=(\ov{\pa}+B)(\pa+A)+2V=(\pa+A)(\ov{\pa}+B)+2U,\tag1
$$
where $A$, $B$, $V$, and $U$ are functions of $(x,y)$, $z=x+iy$. It is
characterized by the {\it electric potential} $V(x,y)$ and the {\it
magnetic field\/} (directed along the third axis):
$$
2H(x,y)=B_z-A_{\ov{z}},\qquad U=V+H\tag2
$$
up to gauge transformations
$$
L\to e^fLe^{-f},\qquad\psi\to e^f\psi,
$$
where $f(x,y)$ is a smooth function and $\psi\in\Cal L_2(\Bbb R^2)$.

We shall consider the periodic case, when the group of periods is
given on the plane, this group being generated by the basis vectors
$\overarrow{T}_1$ and $\overarrow{T}\!_2$, so that the magnetic field
and the electric potential are doubly periodic:
$$
\align
H(\vec{x}+\overarrow{T}\!_1)&=H(\vec{x})=
H(\vec{x}+\overarrow{T}\!_2),\qquad\vec{x}=(x,y),\\
V(\vec{x}+\overarrow{T}\!_1)&=V(\vec{x})=0
V(\vec{x}+\overarrow{T}\!_2).
\endalign
$$

Denote by $K$ the elementary cell in the $(x,y)$-plane, i.e., the
interior of the parallelogram with vertices
$$
K=\{0,0+\overarrow{T}\!_1,0+\overarrow{T}\!_1+\overarrow{T}\!_2\}.
$$

By the {\it magnetic flux\/} we mean the expression:
$$
[H]=\iint_KH(x,y)\,dx\,dy.
$$

\definition{Definitions} a) The case in which $[H]=0$ is called {\it
topologically trivial}. Here all the coefficients of the operator $L$
are doubly-periodic.

b) The case in which $[H]\ne0$ is called {\it topologically
nontrivial}. The coefficients of the operator $L$ are not
doubly-periodic functions on the $(x,y)$-plane. Special attention
will be paid to the case in which $[H]=2\pi n$, where $n$ is an
integer. We shall say that this is the {\it integer case}.
\enddefinition

Suppose, for the sake of simplicity, that the lattice is rectangular,
$$
\overarrow{T}\!_1=(T\!_1,0),\qquad\overarrow{T}\!_2=
(0,T\!_2).
$$

\definition{Definition} The {\it magnetic translations\/} are the following
transformations:
$$
\aligned
\wh T_1\psi(x,y)&=\exp\>\{i f_1\}\>\psi(x+T_1,y),\\
\wh T_2\psi(x,y)&=\exp\>\{i f_2\}\>\psi(x+T_2,y),
\endaligned
\tag3
$$
where
$$
\alignat2
A(x+T_1,y)-A(x,y)&=i\pa f_1,&\qquad
B(x+T_1,y)-B(x,y)&=i\ov{\pa}f_1,\\
A(x,y+T_2)-A(x,y)&=i\pa f_2,&\qquad
B(x,y+T_2)-B(x,y)&=i\ov{\pa}f_2.
\endalignat
$$
\enddefinition

It is known that magnetic translations commute with the operators
$L$, i.e., $L\wh T_i=\wh T_i L$. They also satisfy the
following commutation relation:
$$
\wh T_2\wh T_1=e^{i[H]}\wh T_1\wh T_2, \tag4
$$
where $[H]$ is the magnetic flux through the elementary cell $K$.
These properties can be verified by an elementary substitution.

Thus relation (4) implies that only in the integer case $[H]=2\pi n$
can we define the {\it magnetic Bloch functions\/}:
$$
\wh T_k\psi(\vec{x},\vec{p}\>)=\exp\(ip_k T_k\)\psi
(\vec{x},\vec{p}\>),\qquad k=1,2. \tag5
$$

Here the magnetic quasi-momenta $(p_1,p_2)$ range over the points of
the two-dimensional torus $\vec{p}\in T^2_*$ corresponding to
the reciprocal lattice $K_*$. For fixed quasi-momentum $(p_1,p_2)$, the
operator $L$ acts as an elliptic operator on the sections of the
complex linear bundle over the torus $T^2_*$ with
the connection whose curvature
$H(x,y)$ is determined by formula (5). If the magnetic field,
the electric potential, and the quasi-momenta $(p_1,p_2)$ are real,
then this operator is semibounded, selfadjoint, and has a discrete
spectrum of finite multiplicity
$$
L\psi_j=-\e_j\psi_j,\quad\e_j=\e_j(p_1,p_2),\qquad j=0,1,2,\dots,\;
\e_0\le\e_1\le\e_2 \le\dots,
$$
for given $(p_1,p_2)$. If the energy is large
(\<$\e\to\infty$), then the sizes of the {\it magnetic zones}
$$
\Delta_j=\max_{\vec{p}\in T^2_*}\e_j(p)
-\min_{\vec{p}\in T^2_*}\e_j(p)
$$
tend to zero: $\Delta_j\to0$ as $j\to\infty$. It is assumed that the
magnetic flux is nonzero: $[H]\ne0$.

It is of interest to study the topology of generic operators,
initiated in \cite{N81} (also see the survey \cite{N83}).

\definition{Definition} A selfadjoint operator $L$ with
smooth doubly-periodic $(V,H)$, $[H]=2\pi n$ is called {\it generic\/}
if for all $j$ and all $(p_1,p_2)\in T^2_*$ we have
$$
\e_j(p_1,p_2)\ne\e_k(p_1,p_2),\qquad j\ne k.
$$
\enddefinition

In this case the eigenfunctions $L\psi_j=\e_j\psi_j$ in the
Hilbert space $\Cal H$ of sections of the bundle over the
$(x,y)$-torus $T^2$ depend on the quasi-momenta $(p_1,p_2)\in T^2_*$
as on parameters. This generates a linear complex bundle $\eta_j$
over the torus $T^2_*$, depending on the number $j$. The
corresponding Chern classes $c_1(\eta_j)$, whose study was initiated
in \cite{N81, N83}, as was later established by physicists \cite{T},
play a fundamental role in the theory of the integer quantum Hall
effect, discovered in 1981. These classes may assume different 
random series of values.

\remark{Problem} Study the asymptotic properties of the
integers $c_1(\eta_j)$ as $j\to\infty$ for generic operators $L$ when
$[H]=2\pi n$.
\endremark

The eigenfunction $\psi_j(x,y,p_1,p_2)$, where
$L\psi_j=\e_j(p_1,p_2\)\psi_j$, in the generic case possesses
a manifold of zeros
$$
\psi_j(x,y,p_1,p_2)=0.
$$
This equation expresses two real conditions. Hence, generally
speaking, the manifold of zeros $M_j$ is a two-dimensional real
submanifold in the Cartesian product of tori
$$
M_j\subset T^2\times T^2_*,\qquad M_j=\wt M_j\cup
\skew3\wt{\wt M}_j,
$$
where $\skew3\wt{\wt M}_j$ is projected into a finite number of
points of the torus $T^2_*$, while $\wt M_j$ at a generic
point is given locally in the form of a graph
$$
p_1=F_{j_1}(x,y),\qquad p_2=F_{j_2}(x,y).
$$

The problem of studying the analytic properties of the collection of
vector functions $\vec{p}=\vec{F}_j(x,y)$
was posed in \cite{N83}. It seems probable that the
problem of the spectral theory of the operator $L$ can be effectively
solved on the basis of these notions.

Of special interest is the nonrelativistic Pauli operator for spin
$1/2$ on the plane with an electric field parallel to the plane and a
magnetic field perpendicular to it. This is a vector operator. It has
the form (the charge, mass, and Plank constant are assumed equal to $1$)
$$
P=\sum_{\alpha=1}^2(\pa_{\alpha}-iA_{\alpha})^2+ (W+\sigma_3H),\tag6
$$
where $x_1=x$, $x_2=y$,
$\pa_{\alpha}=\pa/\pa_{\alpha}$, while $(A_1,A_2)$ is
the vector potential, $W$ is the electric potential,
$H=\pa_1A_2-\pa_2A_1$ is the magnetic field, and
$$
\sigma_3=\pmatrix1&0\\
0&-1\endpmatrix.
$$
The operator $P$ has the form of a direct sum of two scalar operators
$$
\gathered
P=L_+\oplus L_-,\\
L_+=\sum^2_{\alpha=1}(\pa_\alpha-iA_\alpha)^2+(W+H),\qquad
L_-=\sum^2_{\alpha=1}(\pa_\alpha-iA_\alpha)^2+(W-H).
\endaligned
\tag7
$$

For a zero electric field $W=0$, the Pauli operator may be factorized as
$$
\aligned
L_+&=(\pa+A)(\ov{\pa}+B)=(\ov{\pa}+B)(\pa+A)+2H,\\
L_-&=(\ov{\pa}+B)(\pa+A)=(\pa+A)(\ov{\pa}+B)-2H.
\endaligned
\tag8
$$

It is convenient to choose a ``real'' gauge $B=-\ov{A}$,
$A=A_2+iA_1$. We shall also impose the usual {\it Lorentz condition}
$\pa_1A_1+\pa_2 A_2=0$ or
$$
\op{Im}\(\ov{\pa}A)=0.\tag9
$$

According to the results of \cite{AC, DN80}, in the absence of any
electric field, the Pauli operator $(-P)$, which is always
nonnegative under these conditions, has the point $\e=0$ as
its ground state for rapidly decaying fields $H$ (see \cite{AC})
and doubly-periodic $H$ when $[H]\ne0$ (see \cite{DN80}). Here if
$[H]>0$, then the ground state corresponds to the $+$ sector and
satisfies the first order equation
$$
(\ov{\pa}+B\)\psi=0.
$$
The equation $L_-\psi=0$ in this case has no solutions with the
necessary boundary conditions. If $[H]=2\pi n$, $n>0$, then the basis
of the magnetic Bloch states may be written in the form
$$
\psi=e^\varphi\prod^n_{j=1}\sigma(z-a_j\)e^{az},\qquad
a_j\in\ov{K},
$$
where
$$
\Delta\varphi=-H,\quad\varphi=\frac{1}{\pi}\iint_K\ln|\sigma
(z-z')|\>H(z')\,dx\,dy,\quad z=x+iy;
$$
here $a_1,\dots,a_n$ are arbitrary constants, while $a$ is determined
by them from the condition requiring the quasi-momenta to be real
(see \cite{DN80}).

The degree of degeneracy for the level $\e=0$ is the same as
that for the Landau level for $H=\op{const}$. The magnetic
field enters into the formula only via $\varphi$ and $a$.

\head\S3.\enspace Laplace transformations. Main results\\
(the topologically trivial case) \endhead

For the Schr\"odinger operator $L=(\ov\pa+B)(\pa+A)+2V$, we
start with an arbitrary solution of $L\psi=0$. Then the function
$\wt{\psi}=(\pa+A\)\psi$ satisfies the equation
$$
[V(\pa+A\)V^{-1}(\ov{\pa}+B)+2V]\>\wt{\psi}=0.
$$
Thus we see that if $L\psi=0$, then $\wt L\wt{\psi}=0$,
where
$$
\wt L=V(\pa+A)V^{-1}(\ov{\pa}+B)+2V,\qquad
\wt{\psi}=(\pa+A\)\psi.
$$
Since $V(\pa+A)V^{-1}=\pa+\wt A$, where $\wt A=A-(\ln V)_z$, we
have $\wt L=(\ov{\pa}+\wt B)(\pa+\nomathbreak\wt A)+2\wt V$,
where $\wt B=B$, $\wt A=A-(\ln V)_z$,
$$
\cases\wt H=H+\frac12\pa\ov{\pa}\(\ln V),\\
\wt V=V+\wt H.
\endcases\tag10
$$

It is essential that the final formulas (10) provide the
transformations only in terms of physical quantities.

\proclaim{Lemma 1} Suppose that the potential $V(x,y)$ and the
magnetic field $H(x,y)$ are real and doubly-periodic, and $V\ne0$.
In this case the magnetic fluxes $[H]$ and $[\wt H]$ coincide, while
for the potentials we have the relation
$$
[\wt V]=[V]+[H].
$$
\endproclaim

The proof follows from formula (10) for the Laplace transformations
because $[\pa\ov{\pa}(\ln V)]=0$.

\proclaim {Lemma 2} Let the potential $V$ and the magnetic field $H$
be real, while the operator $L$ is given in the real Lorentz gauge
$$
L=(\ov{\pa}+B)(\pa+A)+2V,\qquad B=-\ov{A},\qquad
\op{Im}A_{\ov{z}}=0.
$$
Then for any solution of $L\psi=0$ the function
$\skew3\wt{\wt{\psi}}=e^{-Q}(\pa+A\)\psi$ satisfies the equation
$\skew2\wt{\wt{L}}\skew3\wt{\wt{\psi}}=0$, where
$\skew2\wt{\wt{L}}=(\ov{\pa}+\skew3\wt{\wt{B}})
(\pa+\skew4\wt{\wt{A}})+2\skew2\wt{\wt{V}}$, and the operator
$\skew2\wt{\wt{L}}$ is gauge equivalent to the image of the
Laplace transformation
$$
\skew2\wt{\wt{L}}=e^{-Q}\wt L\,e^{+Q},\qquad
\skew2\wt{\wt{V}}=\wt V,\quad
\skew3\wt{\wt{B}}=\wt B+Q_{\ov{z}},\quad
\skew4\wt{\wt{A}}=\wt A+Q_z.
$$
Now if $Q=\ln\sqrt{V}$, then the new operator $\skew2\wt{\wt{L}}$
can be expressed in the real Lorentz gauge as follows
$$
\skew3\wt{\wt{B}}=B+\tfrac{1}{2}(\ln V)_{\ov{z}},\quad
\skew4\wt{\wt{A}}=A-\tfrac{1}{2}(\ln V)_z,\quad
-\skew3\wt{\wt{B}}=\ov{\skew4\wt{\wt{A}}},\quad
\op{Im}\skew4\wt{\wt{A}}_{\ov{z}}=0. \tag11
$$
\endproclaim

The proof of this lemma also follows from the formulas specifying the
Laplace transformations.

Speaking of the Laplace transformation for magnetic fields and
potentials, we shall not distinguish, as a rule, operators from the
same gauge equivalence class. But if we are concerned specifically
with real selfadjoint operators, then by Laplace transformations we
mean the transformations $L\to\skew2\wt{\wt{L}}$ effected in
accordance to formulas (11).

Now let us consider infinite sequences of Laplace transformations:
suppose that $V_j=\exp f_j$, then
$$
e^{f_{j+1}}=e^{f_j}+H_{j+1},\qquad
H_{j+1}=H_j+\tfrac{1}{2}\pa\ov{\pa}f_j,
$$
whence
$$
\tfrac{1}{2}\Delta f_j=e^{f_{j+1}}-2e^{f_j}+e^{f_{j-1}}. \tag12
$$

\proclaim{Lemma 3} After the substitution $f_j=\varphi_j
-\varphi_{j-1}$, the infinite sequence of Laplace transformations
reduces to the well-known {\it two-dimensional Toda lattice}
$$
\tfrac{1}{2}\Delta\varphi_j=e^{\varphi_{j+1}-\varphi_j}-
e^{\varphi_j-\varphi_{j-1}}. \tag13
$$
\endproclaim

Indeed, the substitution $f_j=\varphi_j-\varphi_{j-1}$ in (12) yields
$$
\tfrac{1}{2}\Delta\varphi_j=e^{\varphi_{j+1}-\varphi_j}-
e^{\varphi_j-\varphi_{j-1}}+h(z,\ov{z}),
$$
where $h(z,\ov{z})$ is a function that does not depend on the number
$j$. It is easy to get rid of it by using the change
$\varphi_j\to\varphi_j+\alpha(z,\ov{z})$, where
$({1}/{2})\Delta\alpha=h$.

The two-dimensional Toda lattice together with the representation in
the form of an (L-A)-pair was found by A.~Mikha\u\i lov in \cite{Mik}
and, independently, in \cite{LS, BUL} for finite chains with free ends, as
a generalization of the Liouville equation related to Lie algebras.
It was known to classical geometers (see \cite{Dar, Tzi}) as a
sequence of Laplace transforms. The connection of classical geometric
studies with systems from soliton theory  has been observed by
A.~M.~Vassiliev (unpublished),
but no attempts to use this observation were made. A~consequence of
this comparison is the following

\proclaim{Proposition 1} The algebro-geometric solutions\/
\rom(expressed in terms of $\Theta$-functions\/\rom)
found in\/ \cite{Kr81} for the two-dimensional
Toda lattice generate for any $j$ a two-dimensional
Schr\"odinger operator $L_j$ with quasiperiodic coefficients, which
possesses a Bloch solution $L_j\psi=0$ with appropriate analytic
properties on the same Riemann surface $\Gamma$ of finite genus.
Conversely, suppose that we are given a nonsingular algebraic curve
$\Gamma$ on which two distinguished points $P_+$, $P_-$ with local
parameters $w_+$, $w_-$ are fixed, together with the divisor
$D=P_1+\cdots+P_g$ of degree $g$. If the operator $L=L_0$ corresponds
to the data of the inverse problem
$(\Gamma,P_{\pm},w_{\pm},P_1,\dots,P_g)$, then the operators $L_j$
from the chain of Laplace transformations are obtained from the data
$(\Gamma,P_{\pm},w_{\pm},P_1^j,\dots,P_g^j)$, and we have the
following linear equivalence of divisors
$$
P_1^j+\cdots+P_g^j\sim j\(P_+-P_-)+P_1+\cdots+P_g.\tag14
$$
\endproclaim

Recall (see \cite{DKN}) that in the generic case $P_1,\dots,P_g$ are
poles of order one of the Bloch solution of $L\psi=0$ that do not
depend on $x$, $y$. Near the points $P_{\pm}$ the function $\psi$ has
the asymptotics
$$
\aligned
\psi&\sim c_1(x,y\)e^{z/w_+}(1+\sigma(w_+)),\quad P\sim P_+,\\
\psi&\sim c_2(x,y\)e^{\ov{z}/w_-}(1+\sigma(w_-)),\quad P\sim P_-,
\endaligned\qquad z=x+iy.
\tag15
$$

The function $\psi$ is meromorphic on $\Gamma\setminus(P_+\cup P_- )$
and depends on $x$, $y$ as on parameters. Such a function always
satisfies an equation of the form $L\psi=0$, where
$$
L=(\ov{\pa}+B)(\pa+A)+2V,\qquad
A=-\pa\ln c_2,\quad B=-\ov{\pa}\ln c_1.
$$
Indeed, since $(\pa+A\)c_2=(\ov{\pa}+B\)c_1\equiv0$, the function
$(\ov{\pa}+B)(\pa+A\)\psi$ possesses similar analytic properties
and, therefore, differs from $\psi$ only by a constant factor. It is
easy to see that the function $\wt{\psi}=(\pa+A\)\psi$ has the same
poles $P_1,\dots,P_g$ and asymptotics in $P_\pm$ of the form
$$
\alignedat2
\wt{\psi}&=\wt{c_1}(x,y\)w^{-1}_+e^{z/w_+}(1+\sigma(w_+)),&\qquad
P&\sim P_+,\\
\wt{\psi}&=\wt{c_2}(x,y\)w_-e^{\ov{z}/w_-}(1+\sigma(w_-)),&\qquad
P&\sim P_-.
\endaligned
\tag16
$$

More generally, we have the following

\proclaim{Lemma 4} The function $\psi^{(j)}$, being the result of
$j$ applications of the Laplace transformation to the operator $L$,
besides its poles $P_1,\dots,P_g$, possesses singularities at
the points $P_\pm$ of the form
$$
\alignedat2
\psi^{(j)}&=c^{(j)}_1(x,y\)w^{-j}_+e^{z/w_+}(1+\sigma(w_+)),&\qquad
P&\sim P_+,\\
\psi^{(j)}&=c_2^{(j)}(x,y\)w^j e^{z/w_-}(1+\sigma (w_-)),&\qquad
P&\sim P_-.
\endaligned\tag17
$$
\endproclaim

The proof of Proposition 1 now follows if we compare the result of
Lemma 4 with the construction of the algebraic solutions of the
two-dimensional Toda lattice proposed by I.~M.~Krichever \cite{Kr81}.

Explicit formulas involving $\Theta$-functions for the coefficients
of the corresponding operators $L_j$ can be found in \cite{D81, p.~50}
and \cite{Kr81, p.~77}. They imply, in particular, that the
coefficients are quasi-periodic functions of $x$ and $y$, so that the
operators $L_j$ are topologically trivial.

The solutions of a periodic chain of period $N$,
$\varphi_{n+N}\equiv\varphi_n$, correspond to curves with points
$P_+$, $P_-$ such that $NP_+-NP_-\sim0$, i.e., curves on which there
exists a meromorphic function with a unique pole of $N$\<th order at
$P_+$ and a zero of order $N$ at $P_-$. To such solutions correspond
cyclic chains of Laplace transformations. It turns out that under
certain analytical assumptions all such chains are
algebro-geometric, namely, we have the following result.

\proclaim{Proposition 2} Suppose that a chain of Laplace
transformations is cyclic, $V_{i+N}\equiv V_i$, and all the
potentials $V_i=\exp f_i$ are doubly periodic, smooth, real, and
positive functions. Then for all $n$ the Schr\"odinger operators
$L_n$ are algebro-geometric with respect to the zero energy level
$L_n\psi=0$.
\endproclaim

For $N=1$ from the relation (12) we get $\Delta f_1=0$, which in
the doubly periodic case implies $f_1=\op{const}$. The case
$N=2$ reduces to a similar question for the $\sinh$-Gordon equation
$$
\Delta\varphi+\sinh\varphi=0.
$$

The corresponding result in this case was first obtained by Pinkall
and Sterling \cite{PS}. A.~Bobenko considerably simplified the proof
\cite{B}. The proof of Proposition~1 presented below is a direct
generalization of Bobenko's arguments.

It follows from Lemma 3 that cyclic chains of Laplace transformations
correspond to doubly-periodic solutions of the periodic
Toda lattices (13). It is known
(see \cite{Mik}) that such a chain may be represented as an
$(L\text{-}A)$-pair of the form
$$
[\pa+P,\ov{\pa}+Q]=0,\tag18
$$
where $P$ and $Q$ are the following matrices depending on the
``spectral'' parameter~$\lambda$:
$$
P=\pmatrix\pa\varphi_1&1&\hdots&0\\
\vdots&\vdots&\ddots&\vdots\\
0&0&\hdots&1\\
\lambda&0&\hdots&\pa\varphi_n\endpmatrix,\qquad
Q=\pmatrix0&\hdots&0&\lambda^{-1}e^{\varphi_n-\varphi_1}\\
e^{\varphi_1-\varphi_2}&\hdots&0&0\\
\vdots&\ddots&\vdots&\vdots\\
0&\hdots&e^{\varphi_{n-1}-\varphi_n}&0\endpmatrix.
$$

This allows to determine, in the standard way, an infinite series of
commuting flows
$$
[\pa+P,\pa_i+P_i]=0=[\ov{\pa}+Q,\pa_i+P_i],\tag19
$$
where $\pa_i=\pa/\pa t_i$ and $P_i$ is a matrix polynomially depending
on $\lambda$ (see, for example, \cite{DS}). The derivatives
$\xi^{(i)}=(\pa/\pa t_i)(\varphi)$ satisfy for all $i$ the linearized
system
$$
\tfrac12\Delta\xi^{(i)}_k= \xi^{(i)}_{k+1}e^{\varphi_{k+1}-\varphi_k}-
\xi^{(i)}_k(e^{\varphi_{k+1}-\varphi_k}+ e^{\varphi_k-\varphi_{k-1}})+
\xi^{(i)}_{k-1}e^{\varphi_k-\varphi_{k-1}}. \tag20
$$

This system is an elliptic linear system on a compact manifold (the
torus $T^2$), hence there exists only a finite number of linearly
independent $\xi^{(i)}$. This in its turn implies that any
doubly-periodic solutions of the Toda lattice is stationary for an
appropriate higher flow (19) and by a standard argument (see
\cite{DKN2}) is algebro-geometric. This concludes the proof of
Proposition 2.

\head\S4.\enspace Laplace transformations and\\
topologically nontrivial periodic operators \endhead

As indicated above, in the nonsingular periodic case, the Laplace
transformation
$$
H_{k+1}=H_k+\tfrac{1}{2}\Delta f_k,\quad e^{f_k}=V_k,\qquad
e^{f_{k+1}}=e^{f_k}+H_{k+1}
$$
preserves the magnetic flow $[H_{k+1}]=[H_k]$ and changes the mean
value of the potential
$$
[e^{f_{k+1}}]=[e^{f_k}]+[H_{k+1}].
$$

Thus (nonsingular) cyclic chains with nonzero magnetic flux do not
exist.

\definition{Definition} A {\it semi-cyclic chain\/} of length
$n$ is a chain such that
$$
H_n=H_0,\qquad V_n=V_0+C_n.\tag21
$$
In the nonsingular case we obviously have
$$
C_n=n[H_0].\tag22
$$
Here we assume that all the operators in the chain
$L_0,L_1,\dots,L_n$ have nonsingular coefficients and the invariant
expressions are doubly-periodic
$$
\align
H_n(\vec{x}+T_1)&=H_n(\vec{x}+T_2)=H_n(\vec{x}),\\
V_n(\vec{x}+T_1)&=V_n(\vec{x}+T_2)=V_n(\vec{x}).
\endalign
$$
Here $T_1$, $T_2$ are the basis vectors of the lattice of periods in the
plane, and $\vec{x}=(x_1,x_2)$.
\enddefinition

\example{Example 1} Let $n=1$. From the formulas for the
Laplace transformations, we immediately obtain
$$
H_1=H_0,\quad e^{f_1}=e^{f_0}+H_1,\quad\Delta f_0=0.\tag23
$$
In the nonsingular doubly-periodic case, we have
$$
f_1=\op{const},\quad f_0=\op{const},\quad H_0=\op{const}.
$$
Thus the operator $L_1$ is of the form
$$
L_1=L_0+C_1.\tag24
$$
Here $L_1$ is the Landau operator in a homogeneous magnetic field and
trivial (constant) potential. If we know solutions of $L_0\psi_0=0$
in the Hilbert space $\Cal L_2(\Bbb R^2)$, then the function
$\psi_1=Q_0\psi_0=e^{f_0/2}(\pa+A_0\)\psi_0$ gives the solution
$\psi_1\in\Cal L_2(\Bbb R^2)$. Iterating this procedure, we get
$$
L_n=L_0+nC_1,\qquad\psi_n=Q_{n-1}\cdots Q_0(\psi_0)\in
\Cal L_2(\Bbb R^2).\tag25
$$
In this case we obtain the eigenfunctions of the operator $L_0$ with
eigenvalues $(-nC_1)$ in the space $\Cal L_2(\Bbb R^2)$
$$
L_0\psi_n=(-nC_1\)\psi_n,\qquad n=0,1,2,\dots.\tag26
$$
It is easy to see that
$$
C_1=H_0=[H_0]/K=\op{const},\tag27
$$
where $K$ is the volume of the elementary cell. We know that the
Schr\"odinger operator $(-L_0)$ is semi-bounded. There exists a
constant $E_0$ such that
$$
\langle-L_0\psi,\psi\rangle\ge-E_0\langle\psi,\psi\rangle,
\qquad\psi\in\Cal L_2(\Bbb R^2).
$$
Hence the solution $L_0\psi_0=0$ belonging to $\Cal L_2(\Bbb R^2)$ is
possible only in the case $[H_0]>0$. If $W_0=0$ (or $V=H_0$), i.e., if
$$
L_0=(\pa+A)(\ov{\pa}+B)=(\pa_x+iA_1)^2+(\pa_y+iA_2)^2,
$$
then the solutions serving $\Cal L_2(\Bbb R^2)$ can be found in the
form $(\ov{\pa}+B\)\psi_0=0$ for the real Lorentz gauge
$B=-\ov{A}$, $A_{1x}+A_{2y}=0$. (This will be established
below in connection with the theory of quasi-cyclic chains.) In this
case the spectrum of the operator $(-L_0)$ is discrete and has the
levels
$$
\lambda_n=nC_1=nH_0,\qquad(-L_0)\psi_n=\lambda_n\psi_n.
$$
This is precisely the spectrum of the Landau operator (up to a
translation by the constant $H_0/2$). However, if $[H_0]<0$, then we
must begin with the operator
$$
L'_0=(\ov{\pa}+B)(\pa+A),\qquad V=0
$$
and search for the solution in the form $(\pa+A\)\psi'=0$. The
spectrum is obtained by applying the inverse Laplace transformation
(and its iterations):
$$
\gather
\psi'_1=Q'_0\psi'_0=e^{g_0/2}(\pa+A\)\psi'_0,\qquad g_0=\ln W_0,\\
L'_1=L'_0+C'_1,\quad C'_1=-H_0\quad\lambda_n=-nH_0,\quad
(-L'_0)\psi_n=\lambda_n\psi_n.
\endgather
$$

This concludes Example 1.
\endexample

Our main results are related to the following class of
operators.

\definition{Definition} A {\it quasi-cyclic chain\/} of Laplace
transformations of length $n$ is a chain $L_0,L_1,\dots,L_n$ such that
$$
C_n+H_n=V_n=e^{f_n},\qquad H_0=V_0=e^{f_0}.\tag28
$$
\enddefinition

Suppose further that the magnetic flux is positive and
integer-valued, i.e., $[H_0]=2\pi m>0$.

\example{Example 2} Let $n=2$. Let us write out the
equations that follow from the conditions of semi-cyclicity and
quasi-cyclicity:
$$
\alignat2
\tfrac{1}{2}\Delta g&=-C_2-4e^{a/2}\sinh g,\quad g=f_0-a/2
&\qquad&(\text{semi-cyclic case}),\tag29\\
\tfrac{1}{2}\Delta f_0&=C_2-2e^{f_0},\quad e^{f_2}=2C_2-e^{f_0}
&\qquad&(\text{quasi-cyclic case}).\tag30
\endalignat
$$
\endexample

In both cases we have $C_2=2\>[H_0]>0$. This is especially important in
the case of quasi-cyclic chains. Consider the solutions of these
equations (\<$n=2$) that do not depend on $y$. We obviously have the
following solutions
$$
\alignat2
x=\int\frac{dg}{\sqrt{C_2g+4e^{a/2}\cosh g+C}}
&\qquad&(\text{semi-cyclic case}),\\
x=\int\frac{dg}{\sqrt{2e^g-C_2g+C}}, \quad g=f_0,
&\qquad&(\text{quasi-cyclic case}).
\endalignat
$$

It is clear that for the corresponding values of the constant $C$ in
both cases we obtain wide classes of smooth nonsingular periodic
solutions. Being independent of $y$, the expressions
above determine
doubly-periodic operators with arbitrary period in~$y$. For the
magnetic field we have
$$
\alignat2
&H_0=C_2-2e^{a/2}\sinh(f_0-a/2)
&\qquad&(\text{semi-cyclic case}),\tag31\\
&\aligned
V_2&=2C_2-e^{f_0}=e^{f_2}=H_2+C_2,\\
H_2&=C_2-e^{f_0}=H_1
\endaligned
&\qquad&(\text{quasi-cyclic case}).\tag32
\endalignat
$$

To ensure smoothness and nonsingularity of the operator $L_2$ in the
quasi-cyclic case, we must have $e^{f_0}<C_2$. Solutions satisfying
this condition certainly exist, as can be seen from the formulas.
Since the function $f_0(x)$, as well as the potential and magnetic
field, does not depend on $y$, it follows that we can choose such a
(real) gauge so that
$$
L=(\pa_x+iA_1)^2+(\pa_y+iA_2)^2_H+2V,\quad
A_1\equiv0,\quad A_2(x)=A_2,\quad-A'_2=H(x),
$$
where $L=L_0$ (semi-cyclic chains) and $L=L_2$ (quasi-cyclic chains).
In both cases, after the substitution $\psi=e^{iky}\varphi(x,k)$, we
obtain the following equation in $x$:
$$
\gather
\Lambda_k\varphi_k=0,\\
\Lambda_k=\pa^2_x-(k+A_2(x))^2+(2V(x)-A'_2(x)).\tag33
\endgather
$$

In the Landau case, we have $A_2=Hx$, $V={H}/{2}$. After that the
substitution $x'=x-kH^{-1}$ the operator reduces to the harmonic
oscillator $\Lambda_{k0}=\pa^2_x-H^2x^2$. In the case of a
nonhomogeneous magnetic field we have $H(x)=H$ 
$$
A_2=\ov{H}x+A^0_2(x),\quad\text{where }\,
\ov{H}=\frac{[H]}{T_1T_2};
$$
here $T_1$ is the period in $x$ and $T_2$ is the chosen (arbitrary)
period in $y$, which does appear in the formula anyway, because
$$
[H]=T_2\int^{T_1}_0H(x)\,dx.
$$
The function $A^0_2(x)$ is periodic with period $T_1$. The substitution
$x'=x+k\ov{H}^{-1}$ reduces this equation to the family of equations
$$
\gather
\Lambda_k\varphi_k=0,\\
\Lambda_k=\pa^2_{x'}-{x'}^2{\ov{H}}^2+U
(x'-k{\ov{H}}^{-1})-2k{\ov{H}}A^0_2(x),\\
U=2V(x)-H(x)-(A^0_2(x))^2.
\endgather
$$
The potential $V_2$ for the operator $L_2$ in both cases has the form
$$
V_2=V_0+C_2\quad(\text{semi-cyclic case}),\qquad
V_2=H_2+C_2\quad(\text{quasi-cyclic case}).
$$

The quasi-cyclic operators reduce to a one-dimensional family of {\it
oscillator-like} operators:
$$
\Lambda_k=\pa^2_{x'}-{\ov{H}}^2{x'}^2-2\ov{H}{x'}A^0_2(x)+U(x),\tag34
$$
where $\ov{H}$ is a constant and $A^0_2$, $U$ are periodic
functions (above).

Let us consider the doubly periodic nonsingular case.

\proclaim{Proposition 3} Suppose that the magnetic flux is positive
and integer, the point $O$ belongs to the spectrum of the operator
$L_0$ in $\Cal L_2(\Bbb R^2)$ and all the operators $?????$ of the
semi-cyclic chain are nonsingular, while all the $f_j$,
$j=0,1,\dots,n$, are real and smooth. Then the point $(-C_n)$ also
belongs to the spectrum of the operator $L_0$ in $\Cal L_2(\Bbb R^2)$.
\endproclaim

The proof of Proposition 3 is based on the fact that the Laplace
transformation takes magnetic Bloch functions to magnetic Bloch
functions. This fact is readily verified by substitution and direct
calculation. If $\psi_n\ne0$, then the proposition is proved. Under
its assumptions, if $\psi_n\equiv0$, then it follows that
$Q_{n-1}\psi_{n-1}=0$. Suppose that $j$ satisfies $\psi_j=0$ and
$\psi_{j-1}\ne0$. We have $Q_{j-1}\psi_{j-1}=0$. Thus the operator
$L_{j-1}$ coincides with half of the Pauli operator
$$
L_-=L_{j-1}=(\ov{\pa}+B_{j-1})(\pa+A_{j-1}).
$$
However, we know that the eigenfunctions of the principal state are
possible for $L_-$ here only for negative magnetic flows
$[H_{j-1}]<0$, whereas by assumption we have $[H_{j-1}]=[H_0]>0$.
This contradiction proves Proposition 3.

\proclaim{Proposition 4} Equation {\rm (30)} of the quasi-cyclic
chain of length $n=2$ possesses smooth solutions, essentially
depending on both variables $(x,y)$, doubly-periodic with respect to
the square lattice with any period $T>2\pi C_2^{-1/2}$ sufficiently
close to $2\pi C_2^{-1/2}$. These solutions differ little from
$f_0=\ln\(C_2/2)$.
\endproclaim

The proof of this proposition (necessary to establish that equation
(30) does possess a large class of nontrivial doubly periodic
solutions) was communicated to the authors by S.~Kuksin.

\example{Conjecture} For $n\ge5$, the equation of the
quasi-cyclic chain does not have any nonconstant periodic solutions
close to a constant. The constant is isolated.
\endexample

It seems probable that these equations have no solutions
already in the linear approximation. For $n=2,3,4$ solutions in the
linear approximation can be found.

\proclaim{Theorem} Suppose we are given a quasi-cyclic chain of
length $n$ of operators $L_0,\dots,L_n$ with smooth nonsingular
positive potentials $V_0,\dots,V_{n-1}$ and nonsingular magnetic
fields $H_0,\dots,H_n$ with positive flow $[H_0]=\dots=[H_n]$, the
potentials and fields being doubly-periodic with period lattice
$\overarrow{T}\!_1$, $\overarrow{T}\!_2$. In this case the Schr\"odinger
operator $L_n$ possesses two nondegenerate levels $0$, $+C_n$ each of
which is isomorphic to the Landau level in a homogeneous magnetic
field and zero potential. The eigenfunctions of the levels $0$, $+C_n$
can be computed explicitly.
\endproclaim

\demo{Proof} Let us construct the eigenfunctions of the levels
$0$, $+C_n$. By the quasi-cyclicity condition, we have
$V_n=e^{f_n}=H_n+C_n$. Hence the operator $L=L_n-C_n$ has the form
$$
L=L_n-C_n=(\pa+A_n)(\ov{\pa}+B_n).
$$
Since the flow $[H_n]$ is positive and integer, we can extract the
eigenfunctions of the ground state from the paper
\cite{DN80}
$$
\gathered
(L_n-C_n\)\psi=0,\qquad
\psi=e^{\varphi}\prod\sigma(z-a_1),\dots,\sigma(z-a_m)e^{az},\\
\Delta\varphi_n=-\frac{1}{\pi}\iint_K\ln|\sigma(z-z')|\>H_n
(z')\,d^2z',\quad z'=(x',y'),\quad z=(x,y).
\endgathered
\tag35
$$
Here $K$ is the elementary cell, we do not specify the class $a$ and
constants $a_1,\dots,a_m$ here. The eigenfunctions of the level
$L_n\psi=0$ are obtained by Laplace transformations from the
functions $L_0\psi_0=0$. Indeed, we have
$$
\psi_n=Q_{n-1}\cdots Q_0(\psi_0),\quad\text{where }\,
Q_j=e^{{f_j}/{2}}(\pa+A_j).\tag36
$$
Let us put
$$
\psi_0=e^{\varphi_0}\prod^m_{j=0}\sigma(z-a_{j_0}\)e^{a_0z},\quad
\text{where }\,\Delta\varphi_0
=-\frac{1}{\pi}\iint_K\ln|\sigma(z-z')|\>H_0(z')\,d^2z'.\tag37
$$
The constant $a$ is related to $(a_1,\dots,a_m)$ by formulas
presented in \cite{DN80}. The functions $\psi_0,\psi_1,\dots,\psi_n$
are magnetic Bloch by construction. We already know that the level
$(L_n-C_n\)\psi=0$ is isolated, discrete, and isomorphic to the Landau
level \cite{DN80, N83}. Let us prove this for the level $C_n$. The
subspace of functions (36) obtained is isomorphic to the Landau level
(by construction) and determines a subspace in the Hilbert space
$\Cal L_2(\Bbb R^2)$. Let us prove that there is nothing more at this
level. Let us apply the inverse Laplace transformation to the
operator $L_n$. After $n$ steps we obtain the operator $L_0$, while
from the functions $\psi_n$ we get $\psi_0$. For the level
$L_0\psi_0=0$ we already know that there is nothing more at this
level, except the functions found in \cite{DN80}. If the $n$-fold
Laplace transformation $(\psi_n)\to(\psi_0)$ is an isomorphism, then
the Theorem is proved. If it is not an isomorphism, the inverse
Laplace transformation has a nontrivial kernel. This means, for
$n>1$, that for some number $j$ to the operator $L_j$ corresponds a
function $\varphi_j$ such that $L_j\varphi_j=0$ and
$(\ov{\pa}+B_j\)\varphi_j=0$. Arguing as in the proof of Proposition
3, we see that $L_j=(\pa+A_j )(\ov{\pa}+B_j)$. Thus the subchain
$L_0,\dots,L_j$ is quasi-cyclic of length $j<n$, where the constant
$C_j$ is zero. Having in mind the fact that the magnetic fields are
nonsingular and the relation $f_j =\ln V_j$, we see that
$C_j=j\>[H_0]\ne0$; this contradiction proves the theorem.
\enddemo

\remark{Remark} As the authors were told by Hector de Vega, the
equation $\Delta g=1-e^g$ was obtained as the ``Bogomolmy reduction''
of the Ginzburg--Landau equations for a special value of their
parameter that once divides superconductors of types I and II
(see an explicit quotation to the work of de Vega and Shaposhnik
in \cite{NV3}). 
The
applicability of the Ginzburg--Landau equations at this point is
not clear (possibly it necessitates some corrections, as it was
explained to the authors by experts from the Landau Institute).
However, in this situation the ``physical magnetic field'' differs
from ours by a constant (in such a way that the magnetic flux in the
case of smooth nonsingular solutions is zero). The physical magnetic
flux in the Ginzburg--Landau case entirely consists of
singularities. The study of solutions with isolated singularities is
meaningful in our case also. In this case nontrivial decreasing
solutions as $|z|\to\infty$ (\<$z=x+iy$) are possible in the case
$C_2=0$ as well, where we have an {\it elliptic Liouville equation},
$\Delta g=-e^g$. Its general solution is given by the following
\endremark

\proclaim{Proposition} For an arbitrary analytic function $f(z)$
the function $\varphi$ defined as
$$
e^\varphi=\frac{|f(z)|^2|f'(z)|^2}{(1+|f(z)|^4)^2}\tag38
$$
satisfies the following elliptic version
of Liouville equation
$$
\Delta\varphi+32\>e^\varphi=0.\tag39
$$
\endproclaim

\demo{Proof} It can be checked that the function $\varphi$ of the form
$$
e^\varphi=\frac{r^2}{(1+r^4)^2},
$$
where $r=|z|^2$ satisfies this equation. Now the statement follows from
the following symmetry of Liouville equation: if $\varphi(z,\ov{z})$
is a solution of (39) then $\varphi(f(z),\ov{f(z)})+2\log|f'(z)|$,
where $f(z)$ is analytic, is a solution too.

Taking a rational or elliptic function for $f(z)$ we obtain solutions
with singularities at the poles.
The search for solutions with singularities also makes sense in the
case $C_2\ne0$.
\enddemo

\head Appendix I {\rm(S.~P.~Novikov)}\\
Difference analogs of Laplace transformations\\
and two-dimensional Toda lattices\endhead

In the continuous case, the Laplace transformation $L\to\wt L$,
$\psi\to\wt\psi$ was formally the same in the elliptic and the
hyperbolic case: one could be obtained from the other by the formal
substitution
$$
\gather
\pa\to\pa_x,\qquad\ov{\pa}\to\pa_y,\\
(\ov{\pa}+B)(\pa+A)+2V\to(\pa_x+B)(\pa_y+A)+2V
\endgather
$$
(here, of course, the global statements of the problem are entirely
different).

For difference operators, the hyperbolic and selfadjoint elliptic
case will already differ on the formal level. The separating out of
classes of difference operators that possess Laplace transformations
seems important, since this would indicate the most natural classes
of operators with maximal ``hidden symmetry'' (in some sense) and
therefore with the nicest properties.

\subhead\rom{1.}\quad Hyperbolic case\endsubhead
Let the operator $L$ be of the form \cite{Kr85}
$$
L\psi_n=\psi_n+a_nT_1\psi_n+b_nT_2\psi_n+c_nT_1T_2\psi_n=0, \tag39
$$
where $n=(n_1,n_2)$, $N_j\in\Bbb Z$, and $T_1,T_2$ are the
translations determining the lattice,
$$
T_1\psi_n=\psi_{n+T_1},\quad T_2\psi_n=\psi_{n+T_2}\quad
n+T_1=(n_1+1,n_2),\quad n+T_2=(n_1,n_2+1).
$$

\proclaim{Lemma 5} The operator $L$ has a factorization of the form
$$
L=f_n[(1+u_nT_1)(1+v_nT_2)+w_n].\tag40
$$
\endproclaim

The proof of the lemma is elementary. We come to the following
formulas
$$
f_n=(1+w_n)^{-1}=a_{n-T_1}b_nc^{-1}_{n-T_1},\quad
u_n=u_n=a_n/f_n,\quad v_n=b_n/f_n.\tag41
$$
Note that formulas (41) are local, i.e., to obtain the factorization
we need not solve any difference equations.

\definition{Definitions} The {\it Laplace transformation} is
the assignment $L\to\wt L$, $\psi\to\wt\psi$, where
$$
\gathered
\wt L\wt\psi=0,\qquad L\psi=0,\\
\wt L=\frac{w_n}{1+w_n}[(1+v_nT_2\)w^{-1}_n(1+u_nT_1)+1],\qquad
\wt\psi=(1+v_nT_2\)\psi_n.
\endgathered\tag42
$$

The {\it inverse Laplace transformation} is the assignment
$L\to\skew2\wt{\wt{L}}$, $\psi\to\skew3\wt{\wt{\psi}}$, where
$$
\skew2\wt{\wt{L}}\skew3\wt{\wt{\psi}}=0,\quad
\skew3\wt{\wt{\psi}}_n=(1+q_nT_1\)\psi_n,\quad
L=g_n[(1+p_nT_2)(1+q_nT_1)+s_n].
$$
\enddefinition

\remark{Remark\/ \rm1} Just as in the continuous case, the
Laplace transformation and its inverse are indeed inverse to each
other. This is verified in an obvious way.
\endremark

\remark{Remark\/ \rm2} It is easy to see that to each pair of basis
periods $(T_1^{\pm1},T_2^{\pm1})$ and $(T_2^{\pm1},T_1^{\pm1})$ there
corresponds a Laplace transformation quite similar to (40)--(42),
where we have the pair $(T_1,T_2)$ and $(T_2,T_1)$ for the inverse
transformation. The algebra generated by  all Laplace transformations
is extremely interesting in this case.
\endremark

In the gauge equivalence classes $L\to gLg^{-1}$, the
{\it potential} $w_n$ and the {\it nonphysical magnetic field} $H_n$ are
invariants,
$$
w_n+1=C_{n-T_1}a^{-1}_{n-T_1}b^{-1}_n,\qquad
e^{H_n}=a_nb_{n+T_1}b^{-1}_n a^{-1}_{n+T_2}.\tag43
$$

\proclaim{Lemma 6} The Laplace transformation depends
only on the invariants of the gauge transformation and can be
expressed as
$$
\gathered
e^{\wt H_n}=e^{H_n}w_{n+T_2}w_{n+T_1}w^{-1}_n w^{-1}_{n+T_1+T_2},\\
1+\wt w_{n+T_1}=e^{-\wt H_n}(1+w_{n+T_2}).
\endgathered\tag44
$$
\endproclaim

The proof is a direct verification of the formula.

The infinite chain of Laplace transformations obtained according to
formulas (44) leads to a completely discrete analog of the
two-dimensional Toda lattice for the magnitudes $w^{(k)}_n$, where the
values of $H^{(k)}_n$ are expressed via (44)
$$
e^{H^{(k+1)}}_n=\frac{1+w^{(k)}_{n+T_2}}{1+w^{(k+1)}_{n+T_1}}.
$$
Here $(w^{(k)}_n,H^{(k)}_n)$ is obtained by the Laplace
transformation from $(w^{(k-1)}_n,H^{(k-1)}_n)$, $k\in\Bbb Z$,
$n=(n_1,n_2)$. This system should be compared with \cite{KLWZ}.

\example{Example 3} Let us consider a cyclic chain of
Laplace transformations
$$
H^{(k+m)}_n=H^{(k)}_n,\qquad w^{(k+m)}_n=w^{(k)}_n.
$$

The case $m=1$ is trivial. For $m=2$ we obtain a discrete analog of
the sinh-Gordon equation, and after reduction, we get
$$
\gather
w^{(1)}_n=C(w^{(0)}_n)^{-1},\\
w_{n+T_1+T_2}=w^{-1}_n(C+w_{n+T_1})(C+w_{n+T_2})(1+w_{n+T_1})^{-1}
(1+w_{n+T_2})^{-1}.
\endgather
$$

For $C=1$ the equation degenerates.
\endexample

\subhead\rom{2.}\quad Laplace transformations for selfadjoint real difference
operators\endsubhead
The ordinary discretization of the Schr\"odinger operator, say, on
a square lattice, where at the given point $n=(n_1,n_2)$ one takes
the (weighted) sum of the values of $\psi$ at the same point and
four its nearest
neighbors $(n_1\pm T_1,n_2)$, $(n_1,n_2\pm T_2)$, does not possess
the factorization necessary for the construction of the Laplace
transformation. The latter can be obtained for the operator $L$ if we
use the {\it equilateral triangular lattice}, where each vertex has $6$
nearest neighbors,
$$
|T^{\pm1}_1|=|T^{\pm1}_2|=|(T_1T_2^{-1})^{\pm1}|.
$$

The operator $L$ has the form
$$
L=a_n+b_nT_1+c_nT_2+d_{n+T_1}T_1T^{-1}_2+b_{n-T_1}T^{-1}_1+
c_{n-T_2}T^{-1}_2+d_{T_2+n}T_2T^{-1}_1.\tag45
$$

\proclaim{Lemma 7} An operator of the form {\rm (45)} with real
coefficients possesses the following factorization
$$
L=(x_n+y_nT_1+z_nT_2)(x_n+y_{n-T_1}T^{-1}_1+z_{n-T_2}T^{-1}_2)+w_n.
$$
\endproclaim

The proof is a simple calculation. Note that here, just as in the
hyperbolic case (40), (41), the factorization is given by local
algebraic formulas, which do not require the solutions of difference
equations. We also have the {\it inverse factorization}
$$
L=(x'_n+y'_nT^{-1}_1+z_nT^{-1}_2)(x'_n+y'_{n-T_1}T_1 +z'_{n-T_2}T_2)+w'_n.
$$

To any solution of $L\psi=0$ a new solution is assigned:
$$
\gather
\wt L\wt\psi=0,\qquad\wt L=(x_n+y_{n-T_1}T^{-1}_1
+y_{n-T_2}T^{-1}_2\)w^{-1}_n(x_n+y_nT_1+z_nT_2)+1,\\
\wt\psi_n=(x_n+y_{n-T_1}T^{-1}_1+z_{n-T_2}T^{-1}_2\)\psi_n.
\endgather
$$

We also have an inverse Laplace transformation induced by the inverse
factorization. The two transformations are mutually inverse up to a
gauge transformation.

Here the discrete analog of the two-dimensional Toda lattice, as well
as those of the cyclic, semi-cyclic, and quasicyclic chains arise.
They will be studied in a forthcoming paper.

\remark{Remark} To each pair of neighboring basis periods
of the same length $(T_1,T_2)$, $(T_2,T_1^{-1}T_2)$,
$(T_1^{-1}T_2,T_1^{-1})$, $(T_1^{-1},T_2^{-1})$, $(T_2^{-1},T_1T_2^{-1})$,
and $(T_1,T_2T_1^{-1})$ there
corresponds one Laplace tranformation. The algebra of
these transformations is being studied. 
\endremark

Introducing complex
values for the fields, we also obtain the discretization
of the Schr\"odinger operator in a magnetic field (what we 
considered above is a discretization of real operators only, where
physical magnetic field is equal to zero.)

\head Appendix II {\rm(S.~P.~Novikov, I.~A.~Taimanov)}\\
Difference analogs of the harmonic oscillator\endhead

Let us consider the real selfadjoint second order difference operator
acting in the space $\Cal L_2(\Bbb Z)$, $n\in\Bbb Z$, and given by
$$
\align
L\psi_n&=v_n\psi_n+c_{n-1}\psi_{n-1}+c_n\psi_{n+1},\\
L&=c_{n-1}T^{-1}+c_nT+v\cdot1,
\endalign
$$
where $T\:n\to n+1$ is the translation operator along the lattice.
Here $c_n,v_n\in\Bbb R$, $T^+=T^{-1}$, and $L^+=L$.

\TagsOnLeft

As in the continuous case, one defines the factorization of the
operator $L$:
$$
\align
L+\alpha&=(a_n+b_{n-1}T^{-1})(a_n+b_nT)\quad\text{or}\tag I\\
L+\alpha&=(p_n+q_nT)(p_n+q_{n-1}T^{-1}).\tag II
\endalign
$$

Even without requiring the coefficients to be real, such a
factorization will exist for any $\alpha$. In order to find the
values of $a_n$, $b_n$, $p_n$, and $q_n$ from $v_n$ and $c_n$, we
arrive at the difference analog of the Riccati equation, which
appears when we factorize the continuous Schr\"odinger operator of
the form
$$
-\pa^2_x+u(x)+\alpha=-(\pa_x+v)(\pa_x-v),\qquad v^2+v_x=\alpha+u(x).
$$
In the discrete case we have
$$
\alignat2
v_n+\alpha&=a_n^2+b_{n-1}^2,&\qquad c_n&=a_n b_n,\tag I\\
v_n+\alpha&=p_n^2+q_{n-1}^2,&\qquad c_n&=q_n p_n.\tag II
\endalignat
$$
If all the coefficients $a_n$ and $b_n$ are real, then
$$
L+\alpha=QQ^+,\qquad Q^+=a_n+b_nT.\tag I
$$
Similarly, if all the coefficients $p_n$ and $q_n$ are real,
$$
L+\alpha=RR^+,\qquad R^+=p_n+q_nT^{-1}.\tag II
$$
The Darboux--B\"acklund transformations $B_\alpha$, by definition,
are given by
$$
\align
L\to Q^+Q&=\wt L=B^{(I)}_\alpha L,\tag I\\
L\to R^+R&=\skew2\wt{\wt{L}}=B^{(II)}_\alpha L.\tag II
\endalign
$$
It is obvious that these transformations can be taken to be inverse
to each other after an appropriate factorization:
$$
B^{(II)}_0(B^{(I)}_{\alpha}L)=L+\alpha,\qquad
B^{(I)}_0(B^{(II)}_{\alpha}L)=L+\alpha.
$$
(Recall that in contrast with the two-dimensional case, the
factorization here requires solving a Riccati type equation and is
therefore nonunique.) The cyclic chains, just as in the continuous
case, are determined from the condition:
$$
L_N=B_{\alpha_N}\cdots B_{\alpha_0}L,\qquad
L_j=B_{\alpha_j}L_{j-1},\qquad L_0=L_N.
$$
It is not difficult, by following \cite{VS1}, to prove the following

\proclaim {Proposition 1} Cyclic chains satisfying the condition
$\sum^N_{j=1}\alpha_j=0$ consist of finite-zone difference operators
$L_j$ corresponding to one and the same Riemann surface of genus
$g\le[{N}/{2}]$ \rom(see \cite{DMN, DKN2}\rom).
\endproclaim

More interesting are the chains for which $\sum^N_{j=1}\alpha_j=h>0$.

\TagsOnRight

\subhead The first difference analog of the harmonic oscillator
\endsubhead
Already for $N=1$ nontrivial phenomena arise. Consider the cyclicity
conditions
$$
L=QQ^+-h,\quad\wt L=Q^+Q=B_h L, \quad Q^+Q=QQ^++h,\quad Q^+=a_n+b_n T.
\tag1
$$

From (1) we obtain an equation of Riccati type, which implies
$$
a_n=a=\op{const},\qquad b^2_n-b^2_{n-1}=h\tag2
$$
or $b_n=\sqrt{nh+b^2_0}$. Suppose that further
$$
a=1.\tag3
$$

We come to the following conclusion: {\it the operator $L=QQ^+$ is
not defined in $\Cal L_2(\Bbb Z)$ as a real operator.}

\proclaim{Proposition 2}\footnote{Similar operators were constructed
in \cite{AS90}. However, the analysis of the corresponding formulas
does not reveal if their authors have the same thing
in mind as we do. Apparently our results are new, at least methodically.}
The operator $L=QQ^+$ described by formulas\/
\rom{(1)--(3)} determines a real selfadjoint operator $L$ in the
space $\Cal H_l\subset\Cal L_2(\Bbb Z)$ if and only if $l\in\Bbb Z$
\rom(the {\rm quantization condition)}.

\rom(Here $\psi_n\in\Cal H_l$ if and only if $\psi_n=0$ for
$n\le-b_0^2/h=l$.\rom) The space $\Cal H_l$ is 
isomorphic to $\Cal L_2(\Bbb Z^+)$ with zero boundary condition for
$k=0$.
\endproclaim

\demo{Proof} If the number $l$ is an integer, then
$$
Q^+_{\pm}=(1\pm\sqrt{(n-l)h}\,T).
$$

Setting $n=l+k$, we come to $\Bbb Z^+$. It is readily verified by
substitution that the operators $L_{\pm}=Q_{\pm}Q^+_{\pm}$ take the
space $\Cal H_l$ to itself if and only if $l$ is an integer. We come
to the following operators in $\Cal L_2(\Bbb Z^+)$:
$$
L_{\pm}=Q_{\pm}Q^+_{\pm},\qquad Q^+_{\pm}=(1\pm\sqrt{hk}\,T).
$$

The spectrum of the operator $L_{\pm}=Q_{\pm}Q^+_{\pm}$ in the space
$\Cal L_2(\Bbb Z^+)$ is the same as that of the ordinary harmonic
oscillator. The ground state will be
$$
L_{\pm}\psi^{\pm}_0=0,\qquad Q^+_{\pm}\psi^{\pm}_0=0,\qquad
\psi^{\mp}_{0k}=\cases
(\pm1)^{k-1}/\sqrt{h^{k-1}(k-1)!},&k\ge1,\\
0,&k\le0.\endcases
$$

The higher eigenfunctions have the form
$$
\psi^{\pm}_m=Q^m_\pm\psi^{\pm}_0,\quad
L_{\pm}\psi^{\pm}_m=mh\psi^{\pm}_m,\qquad m\in\Bbb Z^+.
$$
\enddemo

\proclaim{Lemma 1} The eigenfunctions $\psi^{\pm}_m$ of the operator
$L_{\pm}$ have the form
$$
\psi^{\pm}_{mk}=\frac{(\mp1)^{k-1}}{\sqrt{h^{k-1}(k-1)!}}\cdot
P_m(k)\cdot\Theta(k).
$$
Here the $P_j(k)$ are polynomials such that
$$
P_j(k)=(1-h(k-1\)T^{-1})P_{j-1}(k),\quad P_0\equiv1,\quad\text{and}\quad
\Theta(k)=\cases1,&k\ge1,\\
0,&k\le0.\endcases
$$
\endproclaim

The expression $\psi^2_0$ is a Poisson distribution, while the
polynomials $P_j(k)$ are orthogonal with Poisson weight $\psi^2_0(k)$
on $\Bbb Z^+$. Undoubtedly, they are known, although their
relationship with harmonic oscillators and bosonic commutation
relations, most probably, was never discussed.\footnote{These polynomials
are known as the Charlet polynomials.}

\subhead The second difference analog of the harmonic oscillator
\endsubhead
Let $N=1$. Put $L=QQ^+$ and $\wt L=Q^+Q$. We shall introduce a family
of operators $L$ depending on two constants $c,a\in\Bbb R$:
$$
L(c,a)=Q(c,a\)Q^+(c,a),\qquad \wt L(c,a)=Q^+(c,a\)Q(c,a)
$$
so as to have
$$
a^2Q^+(c,a\)Q(c,a)=Q(ca^2,a\)Q^+(ca^2,a)+D,\qquad D=a^2-1. \tag4
$$
Let us put
$$
Q^+=1+ca^n T,\quad Q=1+ca^{n-1}T^{-1},\qquad a\ne0,\;c\ne0.
$$
Consider the transformation $\tau\:n\to1-n$. We have the formula
$$
\tau Q(c,a)=Q^+(c,a^{-1}\)\tau.\tag5
$$
The relations (4)--(5) are extremely interesting. Apparently, they have
not appeared previously.

\proclaim{Theorem} The spectrum of the operator $L(c,a)$ for $\lambda$ 
in the
semi-interval $[0,1)$ in the Hilbert space $\Cal L_2(\Bbb Z)$ has the
form
$$
\alignat3
&1)\;\; a>1,&\qquad\lambda_n&=1-a^{-2n},&\qquad n&\ge0,\\
&2)\;\; a<1,&\qquad\lambda_n&=1-a^{2n},&\qquad n&\ge1.
\endalignat
$$
The eigenfunctions are the following\/\rom:
$$
\alignat2
&1)\;\; a>1,&\qquad
\psi_{0k}(c,a)&=(-1)^kc^{-k}a^{-(k-1)k/2},\qquad k\in\Bbb Z,\\
&&\psi_n(c,a)&=Q(c,a\)Q(ca^2,a)\cdots Q(ca^{2n-2},a\)\psi_0(ca^{2n},a),\\
&&\psi_{nk}(c,a)&=P_n(k,c,a\)\psi_{0k}(c,a),\\
&&&\qquad P_n(k,c,a)=(a^{-2k}-ca^{2n-2})^n,\qquad n\ge1,\\
&2)\quad a<1,&\qquad
\psi_1(c,a)&=\tau\psi_0\Big(\frac{c}{a^2},\frac{1}{a}\Big),\\
&&\psi_n(c,a)&=Q^+\Big(\frac{c}{a^2},a\Big)\cdots
Q^+\Big(\frac{c}{a^{2n-2}},a\Big)\psi_1\Big(\frac{c}{a^{2n-2}},a\Big),
\quad n\ge2.
\endalignat
$$
\endproclaim

\remark{Remarks} 1. For $\lambda\ge1$ the spectrum of the
operators $L(c,a)$ is not known to the authors. We conjecture that it
is continuous.

2. If $a>1$ and $c<0$, then the operator $L$ may be considered in
$\Cal L_2(\Bbb R)$ so that its restriction to the family of lattices
$\delta+\Bbb Z\subset\Bbb R$ yields our family $L(c,a)$:
$$
L_-(m,\gamma)=(1-\gamma^{x-m-1}T^{-1})(1-\gamma^{x-m}T),\qquad
\gamma>0,\;m\in\Bbb Z.
$$
The ground state $\varPhi_0(x)$ (\<$L\varPhi_0(x)=0$) is a
function satisfying
$$
(1-\gamma^{x-m}T\)\varPhi_0(x)=0\quad\text{or}\quad
\varPhi_0(x)=\gamma^{x-m}\varPhi_0(x+1).
$$
It has the form
$$
\varPhi_0(x)=g(x\)\gamma^{x(2m+1-x)/2},
$$
where $g(x)=g(x+1)$ is any function of period $1$.
\endremark

Let us normalize the ground state by defining $g(x)$ from a continuum
of normalizing conditions so that the expression $\varPhi_0^2(x)$ is
a normalized probability distribution on any lattice of the form
$\delta+\Bbb Z\subset\Bbb R$:
$$
g^{-1}(x)=\exp\bigg(\frac{ax(m+1)}{2}-\frac{ax^2}{2}\bigg)
\Theta\bigg[\frac{a(m+1)}{2}-ax\biggm|a\bigg],
$$
where $a=2\ln\gamma>0$ and $\Theta[u\,|\,a]$ is a theta function:
$$
\Theta[u\,|\,a]=\sum_{n\in\Bbb Z}\exp\bigg(-\frac{an^2}{2}+nu\bigg).
$$

It follows from the analytical properties of theta functions
(\cite{BE})
that $g(x)$ is a smooth function with period $1$. The
normed ground state is given by the formula
$$
\varPhi_0(x)=\frac{\gamma^{x(x-1)/2}}{\Theta
[(m+1-2x)\ln\gamma\mid 2\ln\gamma]}.
$$

Thus the cyclic chains are of two types:

1) $L=B_{\alpha_N}\cdots B_{\alpha_1}L=L'$, where $L$ acts
in $\Cal L_2(\Bbb Z^+)$ for an appropriate ``quantization'' of the
parameters;

2) the operator $L'$ coincides with $L$ after multiplication by a
constant and a translation along $x$ in the natural realization in
$\Cal L_2(\Bbb R)$, where the restriction to the lattices
$(\delta+\Bbb Z)\subset\Bbb R$ generates a family of discrete
operators participating in the definition of the cyclic chain
similarly to the case $N=1$ considered above.

\remark{Added in proof} In the paper \cite{SVZ}, the problem was
essentially posed already. However, that paper contains
unmotivated restrictions. For example, the constant
$\delta=\ln|c|/\ln a$ is assumed rational in \cite{SVZ}.
Further, the assertion in \cite{SVZ} according to which
the relation (4) above or the corresponding relation (14)
in \cite{SVZ} and its consequences (15)--(17) ``clearly define a
spectrum generating algebra'' is incorrect. This assertion is certainly
false, for the case in which $\psi_0$ satisfies the equation
$Q^+\psi_0=0$
and growth exponentially. This is indeed the case when $q>1$,
a situation omitted in the papers \cite{AS91, SVZ}. In this
situation the authors have found a different ground state with eigenvalue
$1-q$ (\<$q^{-1}=a^2$) by using the additional symmetry $\tau$ (see
Theorem~2).

Moreover, the assertion that the relations mentioned above
``clearly define a spectrum generating algebra'' is incorrect for
another reason: it gives no information on the spectrum of
the Schr\"odinger operator $L=QQ^+$ with $\lambda\ge1$ in the Hilbert
space $\Cal L_2(\Bbb Z)$. According to our conjecture (see Remark~1
above) this spectrum is continuous and occupies the entire strip
$\lambda\ge1$.
\endremark

\enddocument